\documentclass[fleqn,10pt]{wlscirep}
\usepackage[utf8]{inputenc}
\usepackage[T1]{fontenc}
\usepackage{hyperref}
\usepackage[final]{pdfpages}
\makeatletter
\AtBeginDocument{\let\LS@rot\@undefined}
\makeatother

\title{Machine Learning Partners in Criminal Networks}

\author[1]{Diego D. Lopes}
\author[2,3]{Bruno R. da Cunha}
\author[1]{Alvaro F. Martins}
\author[4]{Sebasti\'an Gon\c{c}alves}
\author[5]{Ervin~K.~Lenzi}
\author[6]{Quentin S. Hanley}
\author[7,8,9,10,$\dagger$]{Matja{\v z} Perc}
\author[1,*]{Haroldo V. Ribeiro}

\affil[1]{Departamento de F\'isica, Universidade Estadual de Maring\'a -- Maring\'a, PR 87020-900, Brazil}

\affil[2]{Rio Grande do Sul Superintendency, Brazilian Federal Police -- Porto Alegre, RS 90160-093, Brazil}
\affil[3]{National Police Academy, Brazilian Federal Police -- Brasília, DF 71559-900, Brazil}

\affil[4]{Instituto de F\'isica, Universidade Federal do Rio Grande do Sul -- Porto Alegre, RS 91501-970, Brazil}

\affil[5]{Departamento de F\'{\i}sica, Universidade Estadual de Ponta Grossa -- Ponta Grossa, PR 84030-900, Brazil}

\affil[6]{School of Science and Technology, Nottingham Trent University, Clifton Lane, Nottingham NG11 8NS, United Kingdom}

\affil[7]{Faculty of Natural Sciences and Mathematics, University of Maribor, Koro{\v s}ka cesta 160, 2000 Maribor, Slovenia}
\affil[8]{Department of Medical Research, China Medical University Hospital, China Medical University, Taichung, Taiwan}
\affil[9]{Alma Mater Europaea, Slovenska ulica 17, 2000 Maribor, Slovenia}
\affil[10]{Complexity Science Hub Vienna, Josefst{\"a}dterstra{\ss}e 39, 1080 Vienna, Austria}

\affil[$\dagger$]{email: matjaz.perc@gmail.com}
\affil[*]{email: hvr@dfi.uem.br}

\begin{abstract}
Recent research has shown that criminal networks have complex organizational structures, but whether this can be used to predict static and dynamic properties of criminal networks remains little explored. Here, by combining graph representation learning and machine learning methods, we show that structural properties of political corruption, police intelligence, and money laundering networks can be used to recover missing criminal partnerships, distinguish among different types of criminal and legal associations, as well as predict the total amount of money exchanged among criminal agents, all with outstanding accuracy. We also show that our approach can anticipate future criminal associations during the dynamic growth of corruption networks with significant accuracy. Thus, similar to evidence found at crime scenes, we conclude that structural patterns of criminal networks carry crucial information about illegal activities, which allows machine learning methods to predict missing information and even anticipate future criminal behavior.
\end{abstract}

\begin{document}
\rfoot{\small\sffamily\bfseries\thepage/10}%

\flushbottom
\maketitle

\thispagestyle{empty}

\section*{Introduction}

Complexity science has only recently started to become popular among researchers working with crime data~\cite{d2015statistical, jusup2022social}. However, many authors are already convinced that complex networks represent an ideal framework to investigate organized crime~\cite{d2015statistical, luna2020corruption, kertesz2021complexity, granados2021corruption, da2021criminofisica}. In line with other research on social systems~\cite{kadushin2012understanding, hou2020survey, jiang2021operator, wu2021visual, waggoner2021uncovering}, complex networks can suitably describe the intricate relations among criminals and reveal the patterns based on which criminal organizations operate. Beyond theoretical explorations, recent articles have empirically demonstrated that these methods can be useful in investigations involving drug trafficking~\cite{duijn2014relative}, political networks~\cite{ribeiro2018dynamical, martins2022universality}, police intelligence networks~\cite{da2018topology}, cartel detection~\cite{wachs2019network}, money laundering~\cite{garcia2020ai}, pedophile rings~\cite{da2020assessing}, and a range of other examples~\cite{calderoni2017communities, colliri2019analyzing, solimine2020political, wachs2021corruption, nicolas2021conspiracy, joseph2021ties}. 

These investigations have also demonstrated that criminal networks exhibit patterns that tie criminal partnerships not only with individual skills, but also with organizational structures that help criminals to optimize, protect and hide their illegal activities. All these regularities and patterns have great potential in helping police investigations, serving as predictive features of future criminal behavior, missing links between individuals, and other properties of unlawful associations. However, there have thus far been very few attempts to use these network patterns to predict static and dynamic properties of criminal networks with machine learning methods~\cite{ribeiro2018dynamical, lim2019situation, calderoni2020robust, qiao2021utilizing}. The paucity of such studies reflects the challenges of obtaining representations for nodes and edges of complex networks that would allow the encoding of structural patterns into vectors to then be used in machine learning algorithms. Obtaining these vector spaces -- an approach known as graph representation learning -- is one of the newest machine learning paradigms that has been developed and is already showing great promise in various applications~\cite{cai2018comprehensive, zhang2020network, chami2021machine, hamilton2022graph}.

Here, our goal is to fill this gap by presenting a comprehensive investigation of political corruption, criminal police intelligence, and criminal financial networks. We rely on the \textit{node2vec}~\cite{grover2016node2vec} method for obtaining vector representations of nodes and edges from these criminal networks, which are then combined with simple machine learning methods in a series of predictive tasks. Our results demonstrate that network properties extracted from \textit{node2vec} are effective in predicting randomly-removed partnerships from criminal networks and recovering missing relationships with accuracy as high as 98\%. Moreover, these vector representations are very effective for distinguishing between criminal, non-criminal, and mixed relationships in criminal police intelligence networks. In addition to being useful in classification tasks, we have also verified that the representations obtained from \textit{node2vec} predict the total amount of money exchanged among agents of a criminal financial network with excellent accuracy. Finally, our investigation shows that one can predict future criminal partners during the growth of political corruption networks.

Our research thus indicates that the underlying patterns of criminal networks carry crucial information about the associations among criminals, allowing us to recover possible missing links and properties of these connections, and even to anticipate future criminal associations. Furthermore, the impressive accuracy and the simplicity of deploying trained machine learning methods allows us to conjecture that our approach is likely to be very helpful in future police intelligence operations.

\section*{Datasets}

Our results are based on four datasets associated with different types of criminal networks. Two of these criminal networks are related to political corruption scandals in Spain and Brazil. The Brazilian data were first used in Ref.~\cite{ribeiro2018dynamical} and the Spanish data were obtained from Ref.~\cite{martins2022universality}. In both networks, nodes represent people involved in political scandals and connections among them indicate individuals engaged at least once in the same corruption case. The Spanish network has 2,695 nodes and 27,545 edges, while the Brazilian network has 404 nodes and 3,549 edges. In addition, we also have information about the growth dynamics of these networks because we know the date of each corruption scandal. As a result, we can reconstruct the growth of these corruption networks by considering corruption scandals occurring up to a given year. The 437 Spanish scandals used in our study occurred between 1989 and 2018, and the 65 Brazilian corruption cases occurred between 1987 and 2014. 

Our third criminal network was obtained from Ref.~\cite{da2018topology} and comprises records of criminal investigations conducted by the Brazilian Federal Police. People involved in this network are criminals or suspected of illegal activities related to federal crimes (drugs and arms trafficking, organized bank robbery, environmental crimes, crimes against elections and financial systems, money laundering, among others), and connections among them indicate individuals involved in the same police investigation or people with personal relationships uncovered during the investigations. This criminal intelligence network has 23,666 nodes and 35,930 edges. For the main component of this network (8,894 nodes and 17,827 edges), we also have information about the type of association between individuals collected by the Brazilian Federal Police. This information is original to our work and classifies the edges among individuals into three types: criminal, mixed, and non-criminal. Criminal edges connect people that are solely related for unlawful purposes; non-criminal edges connect people that do not have a criminal association and may include family or friendship ties; finally, mixed connections represent associations that are both criminal and personal (for instance, two brothers involved in a criminal investigation).

The last dataset used is also original to our study and it is related to a money-laundering investigation conducted by the Brazilian Federal Police from 2008 to 2014. The raw data correspond to bank transactions related to the misappropriation of federal public funds. After being aggregated, this information yields a criminal financial network where nodes represent people or companies, and the connections indicate financial transactions among them regardless of the cash flow direction and amount exchanged.

\section*{Results}

\begin{figure*}[!ht]
  \centering
  \includegraphics[width=1\textwidth, keepaspectratio]{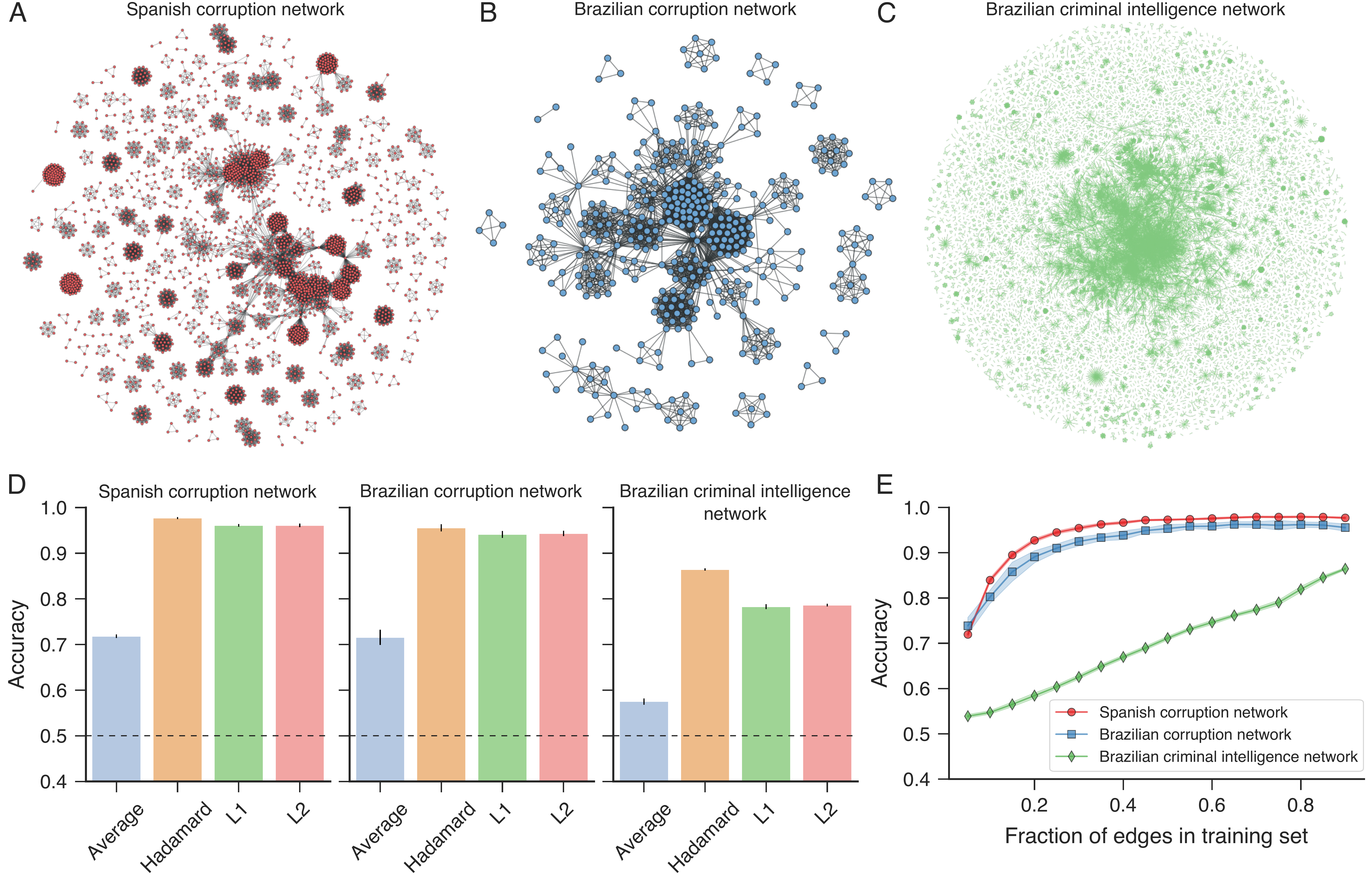}
  \caption{Predicting partnerships in criminal networks. Visualizations of the criminal networks related to (A) Spanish corruption cases, (B) Brazilian corruption cases, and (C) Brazilian criminal intelligence network. In corruption networks, nodes represent people involved in corruption scandals, and connections indicate people participating in the same corruption case. In its turn, nodes in the criminal intelligence network represent people investigated by the Brazilian Federal Police, and an edge between two individuals indicates some co-participation (unlawful or lawful) uncovered by police investigations. (D) Accuracy of logistic classifiers trained for predicting missing links with \textit{node2vec} representations of nodes and different binary operators. The bars stand for the average accuracy estimated from test sets over ten realizations of the embedding and training processes (error bars represent one standard deviation). The test sets are generated by randomly removing 10\% of network edges and sampling the same number of false connections. The horizontal dashed lines represent the baseline accuracy ($0.5$). (E) Accuracy of logistic classifiers as a function of the fraction of nodes in the training set for each criminal network. The markers represent the average accuracy estimated from test sets over ten realizations of the embedding and training processes with the Hadamard operator (shaded regions stand for one standard deviation band). 
  }
  \label{fig:1}
\end{figure*}

We start our investigation by asking whether one can predict criminal partnerships in a static scenario only using structural information of criminal networks. To do so, we consider the final stages (all political scandals) of the Spanish and Brazilian corruption networks and the criminal intelligence network gathered by the Brazilian Federal Police. Figures~\ref{fig:1}A, \ref{fig:1}B, and \ref{fig:1}C depict visualizations of these three networks. We first randomly remove 10\% of the edges of these networks and sample the same number of false connections to create a test set of true and false links. We then use the 90\% remaining edges of these three networks as training sets to fit a logistic classifier~\cite{hastie2016elements} to predict whether the links in the test set are true or false. For training this simple statistical learning method, we generate vector representations of nodes in the training sets using the \textit{node2vec} method~\cite{grover2016node2vec}. This is one of the most popular network embedding methods and consists of finding vector representations that maximize the probability of nodes co-occurring in sequences of biased random walks with fixed lengths. In our analysis, we have fixed the embedding dimension to 256, walk length to 5, number of walks per node to 10, and random walk bias parameters (breadth-first or depth-first) to 1. These choices represent the default setting and make the embedding algorithm similar to \textit{deepwalk}~\cite{perozzi2014deepwalk}. Following Ref.~\cite{grover2016node2vec}, we create vector representations for network edges by combining the vector representation of nodes with four binary operators: average, Hadamard, and L1 and L2 norms. Finally, we associate these vector representations with true edges in the training sets and the same number of randomly sampled false connections.

We thus train the logistic classifiers using these vector representations of true and false edges from the training sets and estimate the accuracy of our approach by calculating the average fraction of correct classifications in the test sets over ten realizations of the train-test split and embedding processes. Figure~\ref{fig:1}D shows these accuracies for the three networks and the four binary operators. The accuracy of the logistic classifiers significantly outperforms the baseline accuracy (50\%) in all cases. Furthermore, in line with the benchmark results presented in Ref.~\cite{grover2016node2vec}, we find the Hadamard operator yields the best performance across our three criminal networks. These best accuracies are remarkably high ($\approx$98\% for the Spanish corruption network, $\approx$96\% for the Brazilian corruption network, and $\approx$87\% for the Brazilian criminal intelligence network), which in turn indicates that structural properties of these networks carry important predictive information about network connections that are well captured by the edge embeddings produced by \textit{node2vec}. 

In Figure~S1, we have compared the performance of \textit{node2vec} with the LINE~\cite{jian2015line} and Mercator~\cite{garc2019mercator} embedding methods. The general accuracies of these other approaches also outperform the baseline accuracy, but are always lower than the scores obtained with \textit{node2vec}. We have also verified how the performance of our approach depends on the fraction of edges used for generating their vector representations. To do so, we have considered only a fraction of edges in the training sets when obtaining the \textit{node2vec} embedding representations and estimated the classification accuracy in the test sets. Figure~\ref{fig:1}E shows these accuracies as a function of the fraction of edges in the training sets used for creating the embedding representations for the three networks. We note that the accuracy in the corruption networks approaches their maximum values much faster than the accuracy in the Brazilian criminal intelligence network. For example, we observe practically no change in the scores of corruption networks after considering $\approx$60\% of edges in the training sets, while the score in the criminal intelligence network monotonically increases with the fraction of edges used in the embedding process. These results indicate that the structure of corruption networks is more redundant than the one observed for the criminal intelligence network. Indeed, corruption networks are formed by a set of complete graphs representing people involved in political scandals that are in turn connected with each other by the recidivism of a small number of agents~\cite{martins2022universality}. In contrast, criminal intelligence networks can have more complex connectivity patterns that are uncovered by police investigations~\cite{da2018topology}.

\begin{figure*}[!ht]
  \centering
  \includegraphics[width=1\textwidth, keepaspectratio]{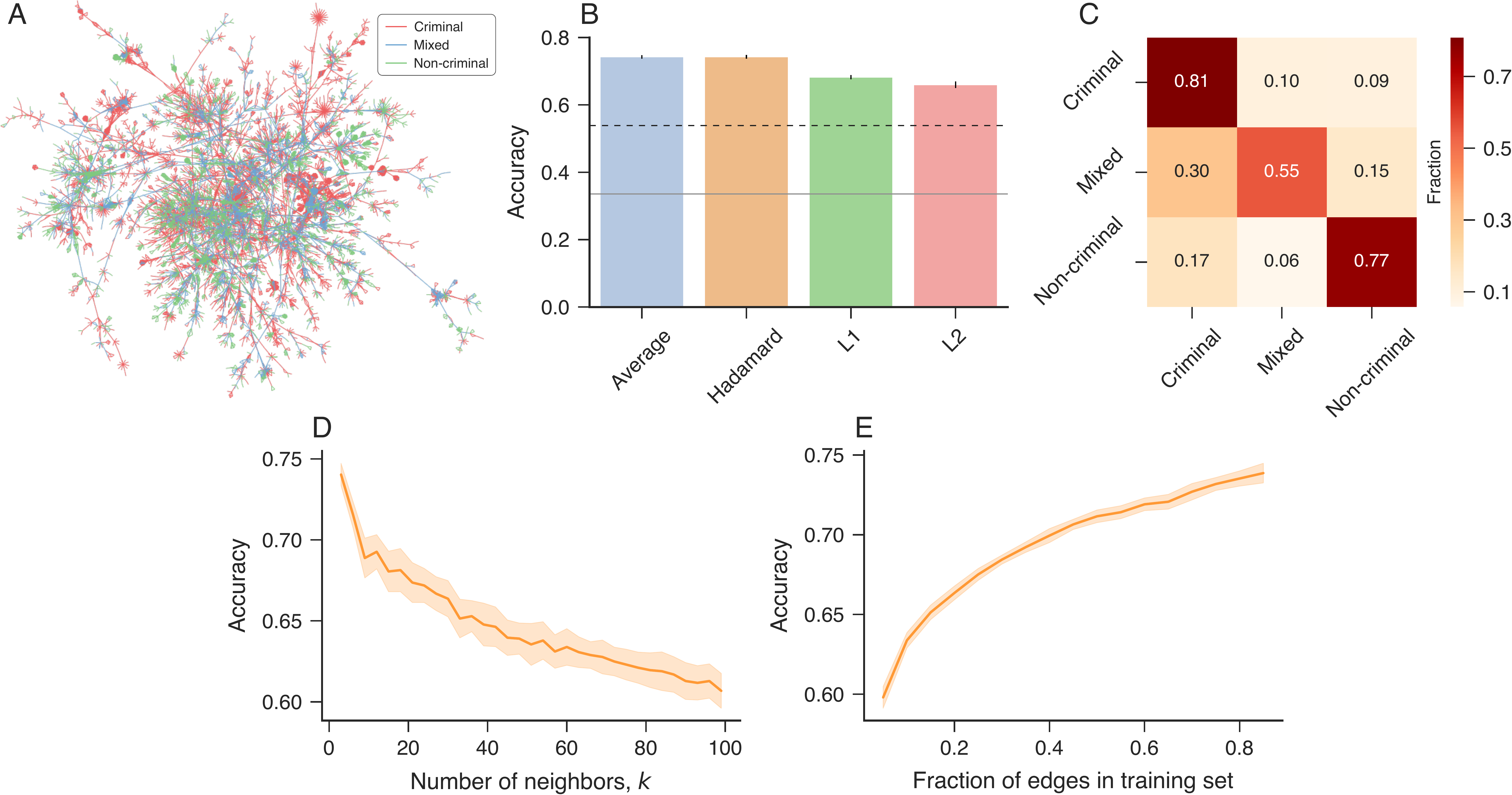}
  \caption{Determining the types of association in criminal networks. (A) Visualization of the three different types of association among people in the giant component of the Brazilian criminal intelligence network. Edges in red, blue, and green represent criminal relationships, mixed relationships, and non-criminal relationships, respectively. (B) Accuracy of $k$-nearest neighbor classifiers ($k$NN with $k=1$) trained with \textit{node2vec} representations and different binary operators. The bars stand for the average accuracy estimated from test sets over ten realizations of the embedding and training processes (error bars represent one standard deviation). The gray continuous line represents the accuracy of a dummy classifier that makes random predictions based on the relative frequency of each type of association in the training set, and the black dashed line indicates the accuracy of a dummy classifier that always predicts the most common type of association in the training set (criminal edge). (C) Confusion matrix associated with the $k$NN classifier predictions (with $k=1$ and the Hadamard operator) for the type of criminal associations in the test sets (rows indicate true labels). (D) Average accuracy in the test sets as a function of the number of neighbors ($k$) in the $k$NN classifiers. (E) Average accuracy in the test sets as a function of the fraction of edges in the training sets. In the last two panels, the solid lines indicate the average accuracy, and the shaded regions stand for one standard deviation band estimated over ten realizations of the embedding and training processes with the Hadamard operator.}
  \label{fig:2}
\end{figure*}

In another application, now focusing on the giant component of the Brazilian criminal intelligence network, we have asked whether the structural properties of this network can be used to determine the type of association among its agents. Figure~\ref{fig:2}A shows a visualization of the giant component of this network where the three types of edges (criminal, mixed, and non-criminal) are depicted in different colors (red, blue, and green, respectively). This time our task is thus to classify the edge types, and to do so, we have again used \textit{node2vec} to generate vector representations of edges by combining the node embeddings with the same four binary operators used in the previous applications. After obtaining the vector representations, we separate (stratified by the three classes of edges) 10\% of data for the test set and use the remaining 90\% as the training set. Furthermore, because the edge classes are imbalanced (54\% criminal, 22\% mixed, and 24\% non-criminal), we have used the random oversampling strategy (randomly replicate minority class examples)~\cite{menardi2014training} to balance the class distribution in the training set. 

We have thus fitted a $k$-nearest neighbors ($k$NN) classifier~\cite{hastie2016elements} to the training data and estimated the average accuracy of the approach in the test set over ten realizations of the embedding process for each binary operator. Figure~\ref{fig:2}B shows these scores in comparison with two dummy classifiers that make predictions based on the relative frequency of each edge type (gray continuous line) and the most frequent edge type (black dashed line). We observe that the accuracy obtained from each binary operator is significantly higher than that of the two baseline classifiers. Again, the Hadamard operator displays the largest accuracy (74\%), followed closely by the average operator. Figure~\ref{fig:2}C presents the confusion matrix of the classification task estimated from the test set using the Hadamard operator (values represent an average over ten realizations of the embedding process). Identifying mixed relationships is more challenging for the $k$NN algorithm as it correctly classifies this edge type in 55\% of cases. In contrast, criminal and non-criminal edges are correctly classified 81\% and 77\% of times, respectively. It is also worth noticing that the algorithm misclassifies mixed relations as criminal edges more frequently than non-criminal ones, which can be regarded as a suitable property when considering that this type of relationship is always related to a possible crime.

We have explored how the number of neighbors ($k$) in the $k$NN classifier affects the accuracy in determining the type of association. Figure~\ref{fig:2}D shows the average accuracy estimated from the test set over ten realizations of the embedding and training processes as a function of the number of neighbors. We observe that the highest scores are obtained for a small number of neighbors and that the accuracy monotonically decreases with the number of neighbors. The results presented in Figures~\ref{fig:2}B and \ref{fig:2}C are for $k=1$ as this value yields the highest accuracy. In addition, we have also verified how the accuracy depends on the fraction of edges used for training the $k$NN model. To do so, we consider a variable fraction of edges ($X$) for training the $k$NN model and use the remaining edges [$(1-X)$\%] as the test set. Figure~\ref{fig:2}E shows that the average accuracy calculated from the test set monotonically increases with the fraction of edges in the training set. However, the accuracy changes are much more intense for lower than higher fractions of edges used for training the learning method. 

\begin{figure*}[!ht]
  \centering
  \includegraphics[width=1\textwidth, keepaspectratio]{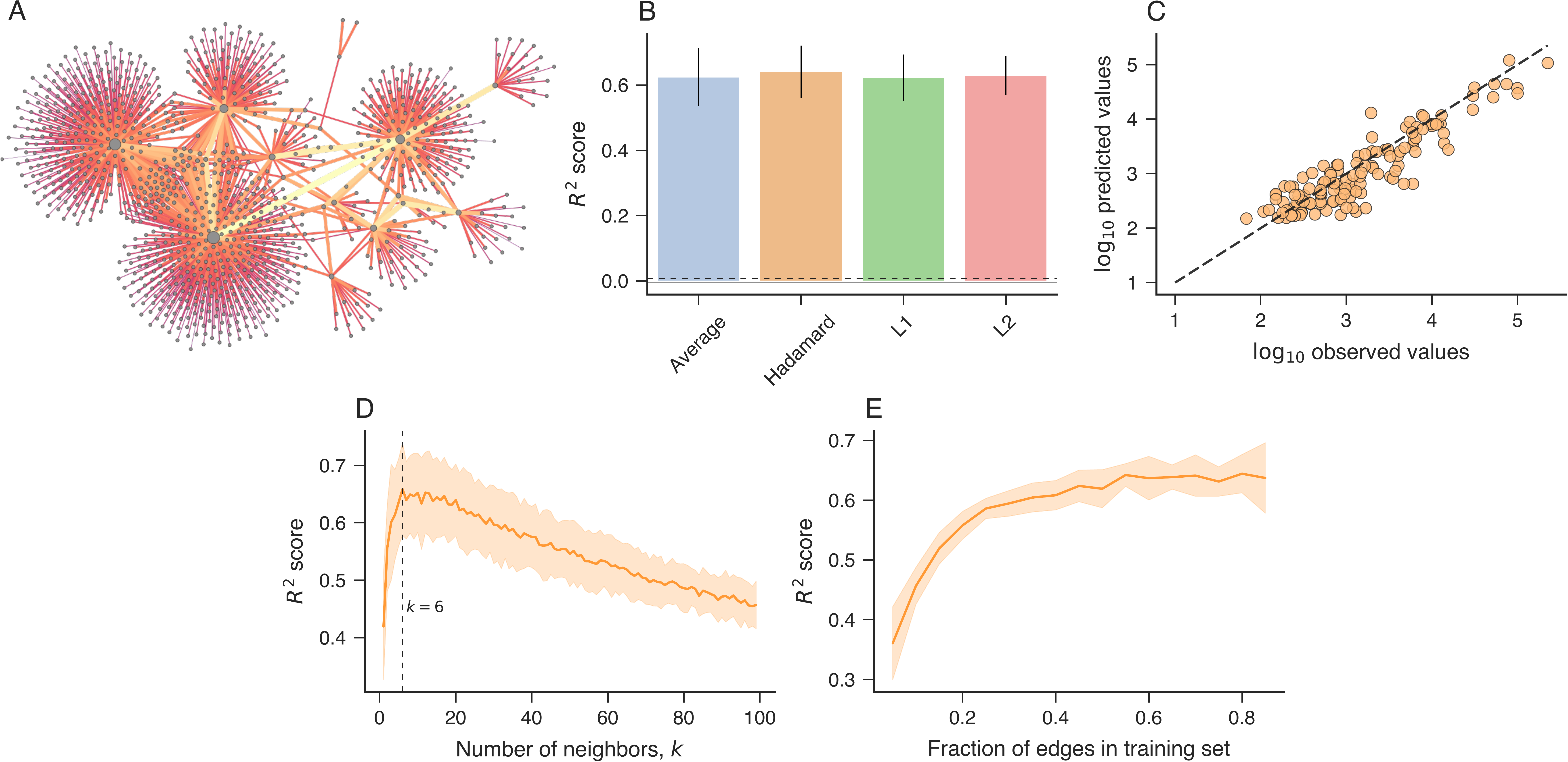}
  \caption{Predicting the total amount of money exchanged among agents of the criminal financial network. (A) Visualization of the criminal financial network. Nodes represent agents (people or companies) and edges indicate financial transactions. The thicker the edge and lighter its color, the larger the amount exchanged between a pair of nodes. (B) Coefficient of determination ($R^2$ score) of the association between the logarithm of the predicted and observed amounts of money exchanged between pairs of nodes in the test sets. These predictions are obtained using $k$-nearest neighbor regressors ($k$NN with $k=6$) trained with \textit{node2vec} representations of edges and different binary operators. The bars stand for the average accuracy and error bars represent one standard deviation over ten realizations of the embedding and training processes. The gray continuous line represents the accuracy of a baseline regressor that always predicts the average value of the training set, and the black dashed line represents the accuracy of another dummy regressor that always predicts the median of the training set. (C) A typical example of the relationship between the base-10 logarithm of the predicted and observed amounts of money exchanged between pairs of nodes in the test sets obtained with a $k$NN regressor ($k=6$) trained with \textit{node2vec} representations of edges and the Hadamard operator. The dashed line represents the 1:1 relationship. (D) Average $R^2$ score as a function of the number of neighbors ($k$) in the $k$NN regressors estimated from the test sets. The vertical dashed line indicates the optimal number of neighbors ($k=6$). (E) Average $R^2$ score on the test sets as a function of the fraction nodes in the training sets. In the last two panels, the solid lines indicate the average $R^2$ score, and the shaded regions stand for one standard deviation band estimated over ten realizations of the embedding and training processes with the Hadamard operator.}
  \label{fig:3}
\end{figure*}

In our third application, we have tried to predict the amount of money exchanged among agents in the criminal financial network only using the structural information of this network. Figure~\ref{fig:3}A depicts a visualization of this network where the edge thicknesses are proportional to the logarithm of the amount of money exchanged between pairs of nodes. Similarly to what we have done before, we have used \textit{node2vec} to create vector representations of all edges in this network with the same four binary operators. However, we do not include any information about the amount of money, such that only the existence or not of (undirected and unweighted) links among nodes is used during the embedding process. After obtaining the vector representations, we have associated them with the logarithm of the amount of money for each network edge and split the resulting dataset into training (90\%) and test (10\%) sets. 

We thus train $k$NN regressors to predict the logarithm of the amount of money and estimate the performance of our approach by calculating the coefficient of determination ($R^2$ score) between the predicted and actual values in the test set. We further average this quantity over ten realizations of the embedding and training processes. Figure~\ref{fig:3}B shows the average $R^2$ score obtained for each binary operator in comparison with two baseline regressors that always predict the average (black dashed line) and median (gray continuous line) of the training sets values. The $k$NN models perform much better than the baselines and yield $R^2$ scores around $0.6$ for all binary operators, but again the Hadamard operator displays the highest performance ($\approx$0.64\%). Figure~\ref{fig:3}C illustrates the typical association between the predicted and observed values in the test set obtained with the Hadamard operator. We have also investigated the roles of the number of neighbors ($k$) and the fraction of edges in the training set ($X$\%) on $R^2$ scores obtained from the test sets [$(1-X)$\%] with the Hadamard operator, as shown in Figures~\ref{fig:3}D and \ref{fig:3}E. We observe that $k=6$ leads to models with the highest performance, and indeed, we have used this value for the results in Figures~\ref{fig:3}B and \ref{fig:3}C. For the fraction of edges in the training set, we note that the $R^2$ score saturates approximately after considering more than 50\% of edges. Although there is certainly room for improving these scores, these results show that our approach works well not only in classification but also in regression tasks.

Finally, in our last application, we have considered the more challenging problem of predicting future criminal partnerships using the structure of criminal networks. We focus on the two corruption networks because we have the network growth dynamics only for these cases. As we have already mentioned, these criminal networks grow by the inclusion of novel corruption scandals containing first-time-offenders and recidivist criminals, with the latter being responsible for creating bonds between different corruption scandals. To approach this problem, we consider scandals occurring up to a given year $Y$ to build the criminal network $G_Y$ and use \textit{node2vec} for creating vector representations for all nodes. We then use these node embeddings to produce vector representations for all network edges and the same number of randomly sampled false connections with the four binary operators. Considering this information as the training set, we train a logistic classifier to distinguish between true and false links. Next, we analyze all corruption scandals occurring after the year $Y$ and collect all connections among nodes already present in $G_Y$. These connections represent future criminal partnerships among agents in $G_Y$. We consider the node embeddings obtained from $G_Y$ to create vector representations for these true future connections and to the same amount of randomly sampled false links that do not occur in the future of $G_Y$, defining our test set. Finally, we apply the trained logistic classifier to determine whether the connections in the test set are true or false and to estimate the average accuracy of our approach over ten realizations of the entire process. Note that no information about scandals occurring after the year $Y$ is used to create the vector representations of edges in the test set or to train the logistic model.

\begin{figure*}[!ht]
  \centering
  \includegraphics[width=1\textwidth, keepaspectratio]{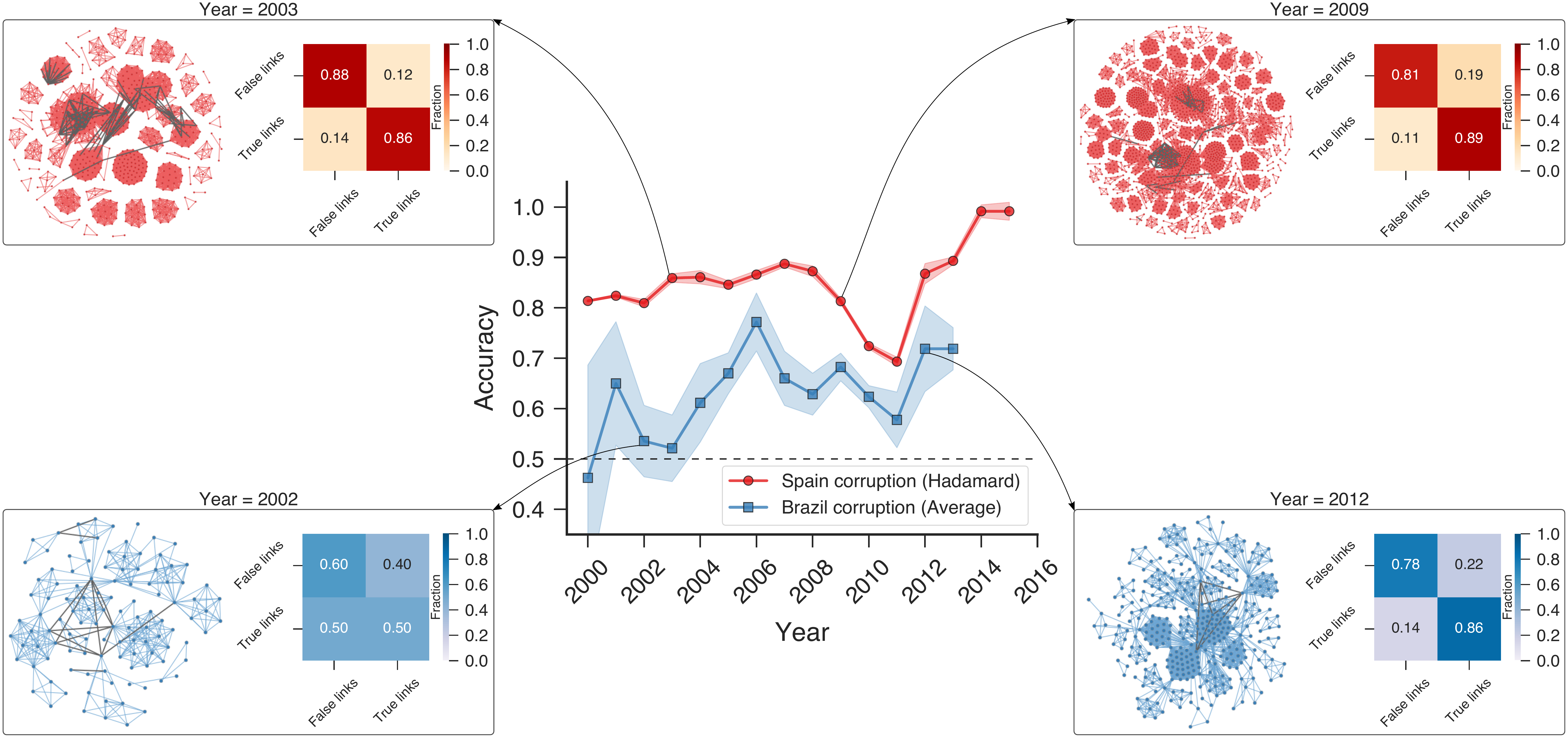}
  \caption{Predicting future partnerships in corruption networks. The central panel shows the accuracy in tasks of predicting future partnerships in the Spanish (red circles) and Brazilian (blue squares) corruption networks created considering scandals occurring up to a given year. The results for the Spanish network use the Hadamard operator, while the ones related to the Brazilian network use the average operator for creating vector representations of edges from the node embeddings obtained with \textit{node2vec}. The test sets of both networks comprise edges among nodes already present in the network that emerge after the threshold year, and the same number of randomly generated false links that do not appear after the threshold year. The markers represent the average accuracy in the test sets estimated over ten realizations of the embedding and training processes (shaded regions stand for one standard deviation band) for different threshold years. The black dashed line indicates the baseline accuracy. The insets depict network visualizations where the colored edges represent connections among nodes that occurred up to the threshold year, while the gray edges represent the links that will appear after the threshold year. These insets also show confusion matrices associated with the tasks of predicting whether future links are true (rows) or false (columns).}
  \label{fig:4}
\end{figure*}

The central panel of Figure~\ref{fig:4} shows the average accuracy in the test sets when considering different threshold years ($Y$) for both the Spanish (red circles) and Brazilian (blue squares) corruption networks. The insets indicated by arrows display visualizations of $G_Y$ for a few years, highlighting future criminal partnerships by gray edges. These insets further show the confusion matrix of the classification process obtained from the test sets. The results in this figure use the Hadamard operator for the Spanish network and the average operator for the Brazilian network because these choices yield the highest average accuracies (see Figures~S2 and S3 for a comparison among the four binary operators and for results obtained with $k$NN classifiers). We observe that the logistic classifiers yield accuracies higher than 0.8 in most years of the Spanish corruption network, significantly outperforming the baseline score ($0.5$). For the Brazilian corruption network, the classification scores do not differ from the baseline accuracy for years before 2003. After this year, the scores fluctuate around $\approx$0.65 and significantly outperform the baseline accuracy. Taken together, these results demonstrate that it is possible to predict future criminal partners using only structural information of criminal networks with good precision. Despite that, the accuracies obtained here are lower than those obtained in our static scenario where edges are removed and then recovered in the final stages of these corruption networks (Figure~\ref{fig:1}A). Thus, link prediction in time-varying networks is indeed more challenging, and results obtained in static scenarios may not generalize well to time-dependent settings.

\section*{Discussion}
We have demonstrated how structural properties of criminal networks and machine learning methods can be used to predict links and link features among actors engaged in nefarious activities. Our research has been carried out using criminal networks associated with political corruption, police intelligence, and financial transactions. In particular, we have shown that simple logistic classifiers trained with embedded representations obtained from \textit{node2vec} are capable of predicting criminal partnerships with excellent precision in static scenarios where a fraction of network edges is removed and then recovered. Beyond predicting whether a link exists or not, we have also shown that $k$-nearest neighbor classifiers trained with vector representations obtained from \textit{node2vec} correctly distinguish between criminal, mixed, and non-criminal relationships in approximately three out of four connections in a police intelligence network. Furthermore, the same embedding approach combined with $k$-nearest neighbor regressors predicts the total amount of money exchanged among agents of a criminal financial network with very good accuracy. Finally, we have shown that structural properties encoded by \textit{node2vec} and learned by simple logistic models can predict future criminal partnerships during the growth process of corruption networks.

Our work, however, does not go without its limitations. One is undoubtedly the information quality used to create criminal networks. Despite the efforts to make such information trustworthy, we must remember these data come from police investigations of illegal and hidden activities, such that missing relationships or noise effects are likely to be present and affect the performance of our machine learning methods. This issue can also partially explain the lower performance we have observed when predicting future criminal associations. Unfortunately, and as also occurs in many other empirical works with social systems, noisy data and missing information are more a rule than an exception. Another limitation is the lack of straightforward interpretations of machine learning methods and the consequent difficulty in deriving causal relationships from these models~\cite{maeda2021black, possati2020algorithmic, le2020remote}. Fortunately, there is a growing consensus that, in addition to delivering high prediction accuracy, machine learning methods must also be capable of producing knowledge from data, a domain that is referred to as ``interpretable machine learning'' and that is experiencing rapid developments~\cite{molnar2020interpretable}, particularly in the context of graph representation learning\cite{li2022survey, hyunju2022providing}.

Despite these limitations, our research strongly corroborates the fact that partnerships among criminals are far from being driven by random circumstances. Indeed, our results indicate that similar to evidence found at crime scenes, criminal associations exhibit patterns and carry crucial information that can be learned by machine learning methods and used to predict missing information or even anticipate the future behavior of agents in criminal networks. Machine learning methods can take vector representations of suspected agents and estimate probabilities for the existence of connections among them and whether they are likely to be criminal or not. It is also worth remarking that we are witnessing a recent surge in research on graph representation learning which in turn yields a large number of techniques for generating effective vector representations for nodes, edges, and entire graphs~\cite{cai2018comprehensive, zhang2020network, chami2021machine, hamilton2022graph}. These methods can be roughly classified into two categories: traditional graph embedding methods and graph neural networks~\cite{khoshraftar2022survey}. The methods we have used are included in the first category, where the vector representations are obtained by optimizing some notion of proximity among nodes of the graph. On the other hand, graph neural networks were proposed even more recently (particularly graph convolutional networks) and belong to the class of deep learning models, where vector representations are obtained by aggregating node neighbors' representations and optimizing loss functions related to specific learning tasks. In addition to being task-specific, graph neural networks can generalize to unseen nodes and explicitly consider node and edge features. Thus, despite the excellent accuracy we have obtained with \textit{node2vec}, exploring other graph representation methods such as graph convolutional networks seems a promising possibility that future research may address. Regardless of being traditional or based on graph neural networks, all these methods can be easily deployed in practical applications involving police intelligence operations, making them potentially useful for helping, guiding, and optimizing police and judicial inquiries.

\bibliography{references.bib}

\section*{Data availability}
Datasets describing the corruption networks and the police intelligence network are freely available on the internet (see Refs.~\cite{ribeiro2018dynamical, martins2022universality, da2018topology}). The dataset for the criminal financial network is available from the corresponding authors upon request. 

\section*{Legal considerations}
The original data on bank transactions and on police intelligence were handled only by Brazilian Federal Police Agents with legal clearance to do so. All data handling was in accordance with the Brazilian law for data protection (Act No. 13709 from 2018), the Brazilian individual rights act (Brazilian Constitution from 1988), the Brazilian Criminal Code (Act No. 2848 from 1940), the Brazilian Criminal Procedure Code (Act No. 3689 from 1941), and the Brazilian Federal Police internal procedures. All data were anonymized before being handed over to the authors.

\section*{Author contributions statement}
D.D.L, B.R.d.C., A.F.M., S.G., E.K.L, Q.S.H., M.P., and H.V.R. designed research, performed research, analyzed data, and wrote the paper.

\section*{Acknowledgements}
We acknowledge the support of the Coordena\c{c}\~ao de Aperfei\c{c}oamento de Pessoal de N\'ivel Superior (CAPES -- PROCAD-SPCF Grant 88881.516220/2020-01), the Conselho Nacional de Desenvolvimento Cient\'ifico e Tecnol\'ogico (CNPq -- Grant 303533/2021-8), and the Slovenian Research Agency (Grants J1-2457 and P1-0403). The authors also thank the Brazilian Federal Police Special Agent Roberto Zaina for providing the bank transaction dataset.

\clearpage
\includepdf[pages=1-3,pagecommand={\thispagestyle{empty}}]{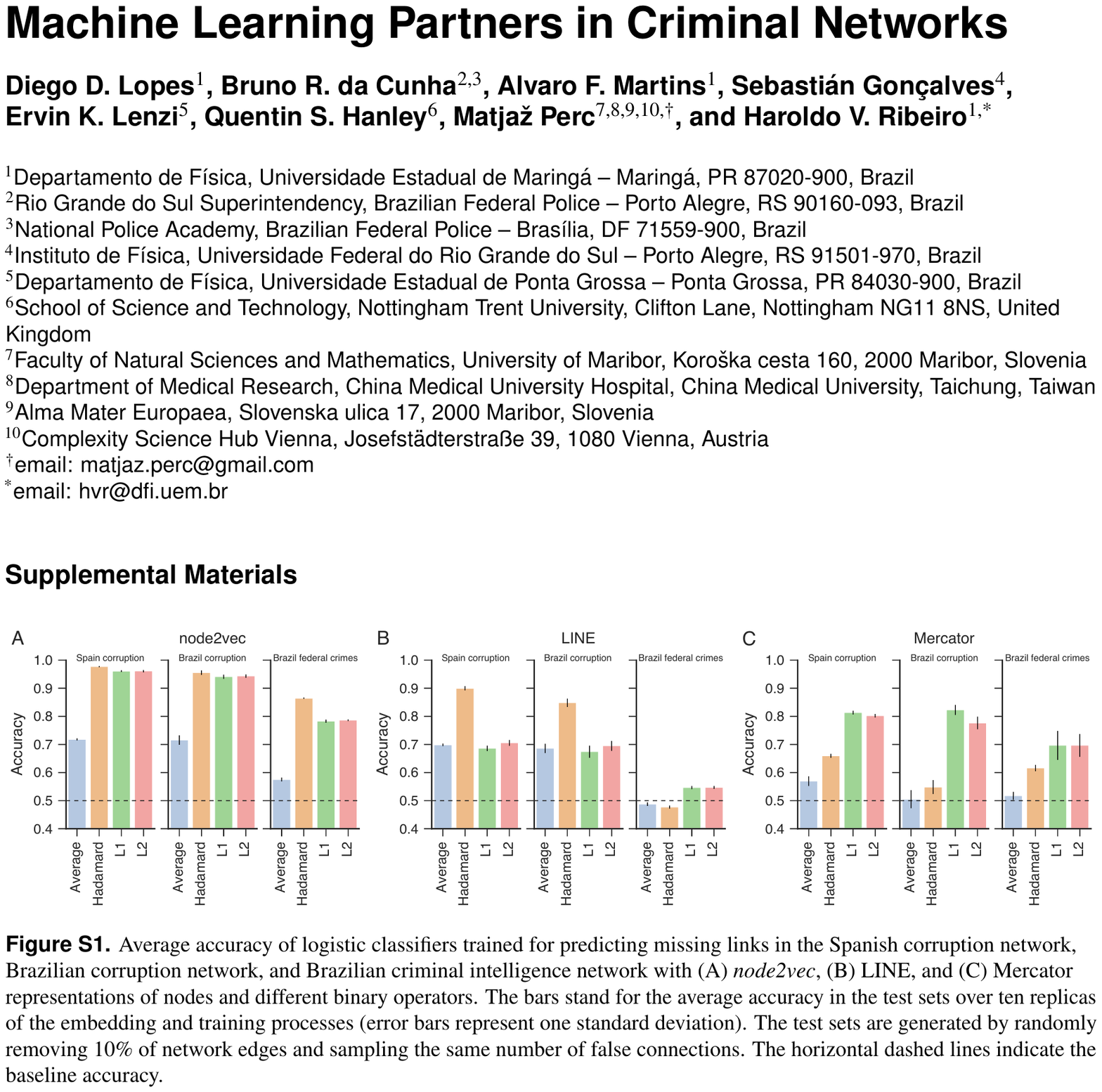}

\end{document}